# *In Vivo* Assessment of Hypertensive Nephrosclerosis Using Ultrasound Localization Microscopy

**Short running title:** Assessment of HN Using ULM

Lanyan Qiu[1,#], Jingke Zhang[2,#], Yi Yang[2,#], Hong Zhang[1], Fu-Feng Lee[2], Qiong He[2,3], Linxue Qian[1,*], Jianwen Luo[2,*]

[1]Department of Ultrasound, Beijing Friendship Hospital, Capital Medical University, Beijing, China

[2]Department of Biomedical Engineering, School of Medicine, Tsinghua University, Beijing, China

[3]Tsinghua-Peking Joint Center for Life Sciences, Tsinghua University, Beijing, China

[#]These authors contributed equally to this work.

## Corresponding author:

Dr. Linxue Qian

Department of Ultrasound, Beijing Friendship Hospital, Capital Medical University, Beijing 100084, China. E-mail: qianlinxue2002@163.com

Dr. Jianwen Luo

Department of Biomedical Engineering, School of Medicine, Tsinghua University, Beijing 100084, China. E-mail: luo_jianwen@tsinghua.edu.cn

**Abstract**

**Purpose:** As a typical chronic kidney disease (CKD), hypertensive nephrosclerosis (HN) is a common syndrome of hypertension, characterized by chronic kidney microvascular damage. Early diagnosis of microvascular damage using conventional ultrasound imaging encounters challenges in sensitivity and specificity owing to the inherent diffraction limit. Ultrasound localization microscopy (ULM) has been developed to obtain microvasculature and microvascular hemodynamics within the kidney, and would be a promising tool for early diagnosis of CKD.

**Methods:** In this study, the advantage of ULM over conventional clinical inspection (serum and urine tests) and ultrasound imaging (Doppler and contrast-enhanced ultrasound imaging) for early diagnosis of HN was investigated. Examinations were carried out on 6 spontaneously hypertensive rats (SHR) and 5 normal Wistar-Kyoto (WKY) rats at the age of 10 weeks.

**Results:** The experimental results showed that the indicators derived from conventional clinical inspection and ultrasound imaging (PSV, EDV, RI, RT, IMAX, mTT and AUC) did not show significant difference between hypertensive and healthy rats ($p > 0.05$), while the TTP derived from CEUS ($p < 0.05$) and the mean blood flow speed in artery of SHR derived from ULM is significantly higher than that of WKY rats ($p < 0.01$).

**Conclusion:** The quantitative results showed that ULM has higher sensitivity than conventional clinical inspection and ultrasound imaging. ULM may promise a reliable solution for early diagnosis of HN.

Keywords: hypertensive nephrosclerosis; microbubble; renal microvasculature;

ultrasound imaging; ultrasound localization microscopy.

## 1. Introduction

Hypertension is a very important public health issue worldwide. According to a study on the prevalence and control of hypertension in 35 to 75-year-old Chinese,[1] nearly half (44.7%) of the Chinese adults suffer from hypertension and just over half (55.8%) of them with stage 2 and above are aware of their diagnosis. Hypertensive nephrosclerosis (HN), as a typical chronic kidney disease (CKD), has a 5% incidence rate in hypertensive population and is the second leading cause of end-stage kidney disease.[2] HN is very difficult to diagnose in the early stage. Once the disease symptoms become apparent, the kidney function is already significantly impaired, and only two treatments including dialysis and kidney transplant are effective at end-stage. Therefore, early diagnosis and assessment of kidney damage caused by hypertension is critical to prevent the progression of HN to kidney failure.[3]

In clinical practice, HN is usually diagnosed by combining the history of hypertension (more than 10 years), progressive proteinuria and slowly progressive chronic renal insufficiency. Proteinuria is a typical symptom of nephropathy, which is caused by various kidney filtration problems. However, there are many diseases associated with the feature of proteinuria.[4] Estimated glomerular filtration rate (eGFR) derived from the measurement of the level of serum creatinine is a common diagnostic parameter of renal function to assess the functional renal unit loss. In the steady state, the serum creatinine level is related to the reciprocal of the level of GFR and can be used to estimate the eGFR. However, the serum level of endogenous filtration markers can also be affected by factors other than the GFR, including tubular secretion or reabsorption, generation, and extra renal elimination of the endogenous filtration marker.[5,6]

There may be divergence between diagnosis based on clinical criteria (such as blood and urine tests) and biopsy pathology.[7] Percutaneous renal biopsy, the gold standard for the diagnosis of HN, is mandatory if an accurate diagnosis is to be made.[8] However, given the potential risk of complications such as bleeding and infection,[9] biopsies are rarely performed for many patients until their late course with secondary hypertension caused by longstanding CKD.[10] This will reduce the diagnostic accuracy of HN and be inimical to early detection of HN. Therefore, a non-invasive and effective methodology for renal function analysis would be an important contribution to early HN diagnosis.

The renal perfusion pressure plays a great role in maintaining renal function.[11] Currently, in clinical practice, noninvasive evaluation of renal perfusion is mainly carried out by renal dynamic imaging based on single photon emission-computed tomography (SPECT).[12] However, this method has poor spatial resolution and cannot observe the dynamic changes of renal cortex and medulla perfusion respectively. Other approaches of medical imaging such as contrast-enhanced computed tomography (CT) and magnetic resonance imaging (MRI) are incompetent in the diagnosis of CKD, because they have toxic side effects on renal function due to the nature of the contrast agent, and their resolution is not high enough to display the arterioles in the kidney.[13]

Ultrasound imaging has become a preferred method for evaluating renal disease because it is noninvasive, real-time, convenient and suitable for multiple follow-ups. To determine the influences of the renal artery stenosis, various indexes are evaluated through conventional B-mode and Doppler imaging, including morphological characteristics, parenchymal echo, peak systolic velocity (PSV), end-diastolic velocity (EDV), resistance index (RI) and other hemodynamic parameters of the large and middle arteries of the kidneys.[14] Contrast-enhanced ultrasound (CEUS) has recently been used as a new imaging technique for quantifying tissue perfusion changes in the kidney.[15] Recent studies have revealed that the CEUS parameters of acute kidney injury,[16,17] CKD,[18,19] and the blood perfusion of kidney transplant derived from perfusion curves in renal cortex are different from those of normal kidney.[20] However, the minimum resolvable distance of conventional ultrasound imaging (generally hundreds of microns) is larger than the diameter of microvascular (< 100 μm) and capillary (< 10 μm) of human body owing to the inherent diffraction limit, which leads to the difficulty of renal hemodynamics assessment. In particular, the low sensitivity of CEUS in the detection of renal arterioles located in the deep renal medulla greatly limits its abilities to evaluate renal blood perfusion.[21,22]

Recently, the resolution of ultrasound imaging has been dramatically increased by using ultrasound localization microscopy (ULM), which is a technique of ultrasound super-resolution imaging.[23,24] By localizing individual injected microbubbles (MBs) and tracking their movements, vascular and velocity maps can be obtained at a scale of micrometers.[25,26] Pathologies which profoundly alter the microvasculature, such as cancer, diabetes and arteriosclerosis, can thus be observed with ultrasound imaging noninvasively.[27–29] ULM has now been applied pre-clinically and clinically to imaging of the microvasculature of the brain, kidney, breast, tumors and other tissues or organs.[30–35]

In this paper, we investigated the advantage of ULM over conventional clinical inspection (serum and urine tests) and ultrasound imaging methods (Doppler and CEUS) for early diagnosis of HN through comparison between the results of 6 spontaneously hypertensive rats (SHR) and 5 normal Wistar-Kyoto (WKY) rats at the age of 10 weeks, when high blood pressure begins to be steady in SHR.[3] In our previous work, the blood flow speed in ischemic-reperfused rat kidney model was found to be much lower than that in normal rats,[32] and the result has been further validated in other studies.[33,34] In order to investigate the relationship between HN and renal blood flow speed, we quantitatively assess the mean blood flow speed and speed gradient of artery and vein on the ULM velocity map, respectively. The conventional clinical inspection indicators and ultrasound imaging indicators were compared with those derived from ULM. Blood pressure and heart rate were recorded, and histological analysis was performed to observe the change of renal micro-structure caused by hypertension.

## 2. Materials and Methods

### a) Animal preparation

SHR is the most commonly used animal model in the study of hypertensive disease,[3,36] and the WKY rats are commonly used as control. Approved by Medical

Ethics Committee of Beijing Friendship Hospital (No.18-1007), SHR and WKY rats at the age of 10 weeks, when the hypertensive symptom began to be stable for SHR,[37] were used in this study. Six SHR ($n=6$) and five WKY rats ($n=5$) are used in the final statistical analysis.

b) **Blood pressure and urine tests**

On the day before the ultrasound imaging experiments, tail arterial systolic, diastolic and heart rate of each rat was measured by a non-invasive blood pressure measurement system for small animals (CODA® Monitor, Kent Scientific, Torrington, CT, USA) without anesthesia. Thereafter, the rats were placed in metabolic cages to collect their urine during 24 hours, from which urine protein was detected.

c) **Conventional ultrasound imaging**

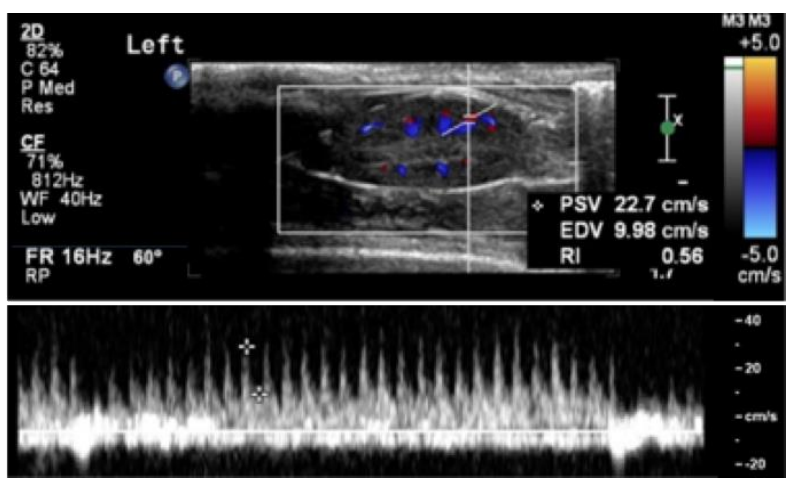

Figure 1. Measurement of PSV, EDV and RI in Doppler imaging mode.

After anesthesia by intraperitoneal injection of 10% chloral hydrate with a dose of 3 ml/kg, the rats' abdomens were shaved to improve ultrasound penetration. Routine B-mode and Doppler ultrasound evaluation were then performed on the rats' kidneys by an iU Elite ultrasound system (Philips, Bothell, WA, USA), equipped with an L12-5 probe. To measure the blood flow speed as close as possible to the cortex, interlobar artery was selected to measure the PSV and EDV using Doppler imaging mode, as shown in Fig. 1. The RI of interlobar artery was computed [RI=(PSV-EDV)/PSV] and recorded as well.

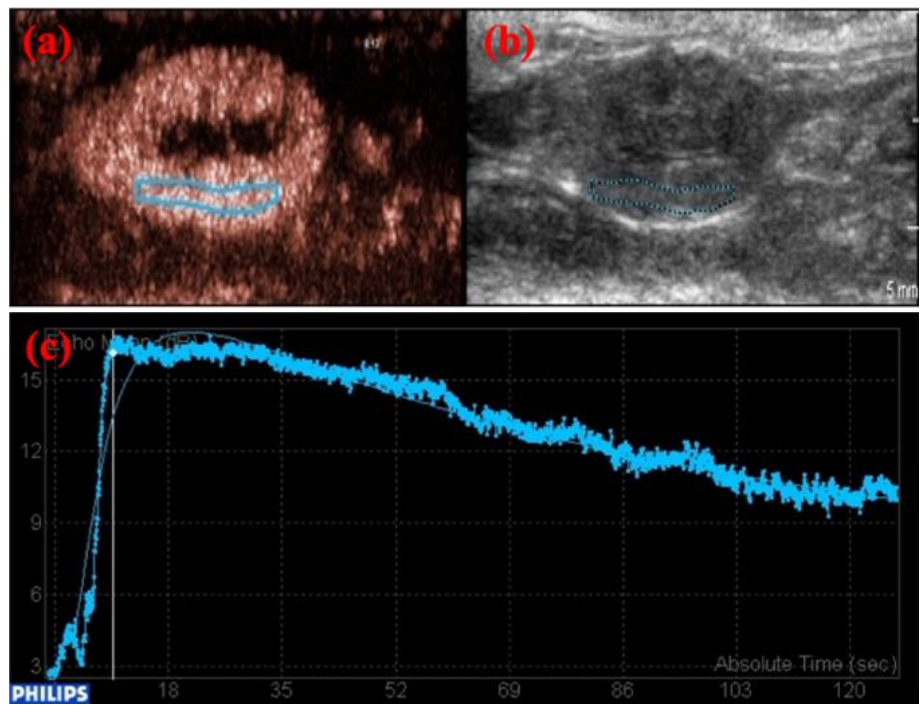

Figure 2. A time-intensity curve (TIC) was obtained from the CEUS renal perfusion video using QLAB software. (a) An image in harmonic imaging mode, where the blue box indicates the ROI used for the calculation of TIC. (b) The corresponding B-mode image. (c) The curves of the original TIC (dotted line) and the fitted TIC using LDRW-WIWO model (thin line).

After that, a bolus of 100 μL MB solution (SonoVue, Bracco, Milan, Italy), diluted with saline by 1:9, was injected to the tail vein of the rats, and flushed with another 200 μL saline. Renal CEUS was performed on the sagittal plane of the left kidney with an L9-3 probe. Both B-mode and contrast-enhanced images were acquired in the CEUS mode, the 1.5 - 2 min video was recorded after bolus injection. The time-intensity curve (TIC) of a ~10 mm$^2$ region of interest (ROI) inside the renal cortex was obtained from the CEUS video using LDRW-WIWO-fitting (local density random walk wash-in wash-out) of QLAB software (QLAB 10.5, Philips, Bothell, WA, USA), as shown in Fig. 2. The quantitative perfusion parameters [38,39] [Fig. 3] including rise time (RT [s]), time to peak (TTP [s]), peak intensity (IMAX [dB]), mean transit time (mTT [s]) and area under the time-intensity curve (AUC, dB·sec) were obtained directly from QLAB.

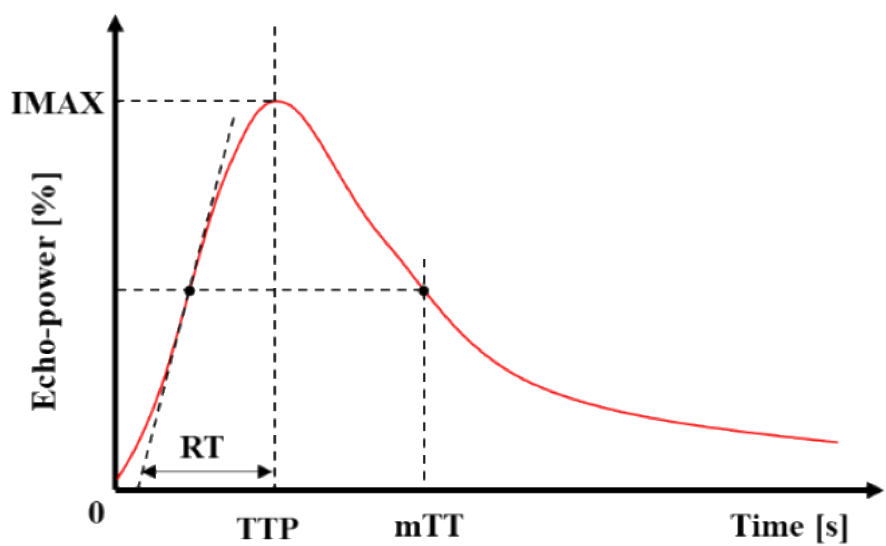

Figure 3. Definitions of quantitative parameters in quantitative analysis of the TIC. IMAX refers to the peak intensity of the TIC curve; TTP is the time when IMAX is reached; RT is the period of time from the intersection point of transverse axes and tangent line of the half maximum point (when the echo-power signal rises to half the peak) to TTP; mTT is the time when the echo-power intensity decreases to half of IMAX; and AUC is the area under the fitted curve.

**d) ULM**

10 seconds after MBs injection, a Vantage 256 system (Verasonics, Kirkland, WA, USA) equipped with an L15-Xtech linear array probe (Vermon, Tours, France) was used to acquire the radiofrequency (RF) channel data on the maximum sagittal plane of the left kidney. Each frame was compounded coherently by 5 steered plane-wave images (from -12° to 12°, with a step of 6°). The pulse repetition frequency (PRF) was 2,000 Hz, and the effective frame rate was 400 Hz. A low mechanical index (MI < 0.2) was adopted at the 15 MHz transmission frequency to reduce MB destruction. Finally, 30,000 frames of RF channel data were acquired within 15 seconds. The flowchart of ULM is shown in Fig. 4. Specific methods and parameters are described as follows.

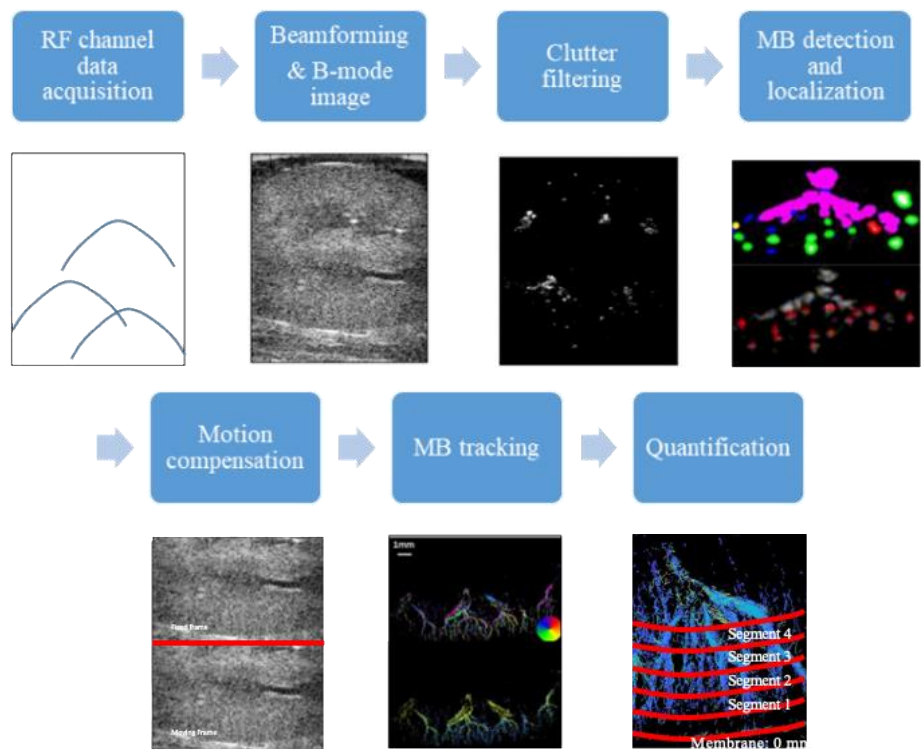

**Figure 4.** Flowchart of ULM. Delay-and-sum beamforming was first performed offline on the ultrasound channel data.[40] Thereafter, singular value decomposition (SVD) based spatiotemporal filtering was applied to extract the MBs from the surrounding tissues.[41]

- 10 -

Respiratory and cardiac motion was estimated on the B-mode images with a 2-D phase correlation-based rigid geometric image registration method.[42] The individual MBs were localized by their intensity-weighted centroids [43] and then motion compensated. Finally, a localization density map with super-resolution was obtained by accumulating the centroids of MBs extracted from all the images. And the blood flow direction and speed were estimated by tracking the movements of MBs.

1) **Clutter filtering**

SVD was performed every 100 frames. The singular value was log-compressed and fitted to a 10$^{th}$ order polynomial. Then the order of SVD filtering was selected with an adaptive method.[44] The singular values below the threshold were set to zeros, and the frames with tissue removal were obtained by using spatiotemporal reorganization.[41]

2) **MB detection and localization**

After a threshold operation to eliminate noise and an interpolation operation in the lateral direction, the individual MBs were screened by morphologic characteristics and localized by their intensity-weighted centroids, as in our previous work.[32]

3) **Motion compensation**

A 2-D phase-correlation-based rigid image registration method was used to estimate and compensate the in-plane tissue motion caused by breathing or heartbeat.[45]

4) **MB tracking**

Individual MBs were paired between each two adjacent frames with a modified Hungarian algorithm.[46] Then the trajectories, in which MBs were not tracked on at least 4 consecutive frames, were rejected. Finally, a Kalman filter was utilized to smooth the MB trajectories and lead to a more precise blood flow speed estimation.[47] The blood flow direction and speed were calculated using the Kalman-filtered trajectories.

5) **Quantification**

After MB tracking, the blood flow speed of renal microvessels can be estimated to obtain the ULM velocity map [Fig. 5(a)]. To investigate the individual changes of the arteries and veins caused by hypertension, the arteries and veins were distinguished according to the moving direction of each MB trajectory. More specifically, the MBs move from the proximal to distal end would be considered as moving in the artery, and vice versa. The arteries and veins were preserved to obtain arterial microvasculature [Fig. 5(b)] and venous microvasculature [Fig. 5(c)], respectively.

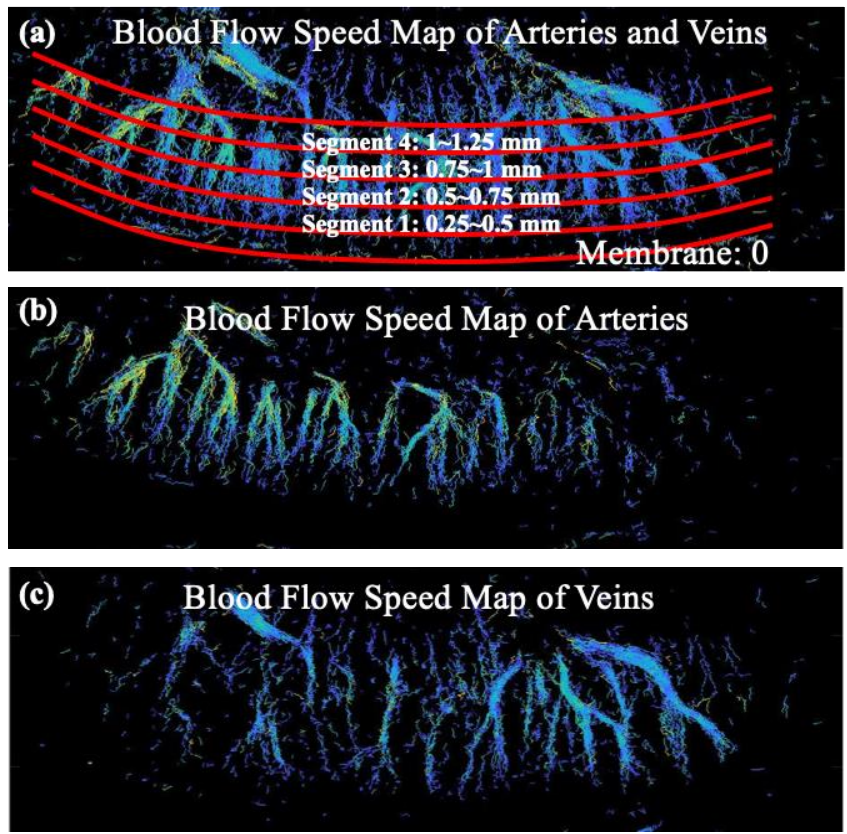

**Figure 5.** (a) The blood flow speed map of arteries and veins are estimated from Kalman-filtered trajectories. The bottom red curve indicates the manually depicted renal

membrane (0 mm). The above five red curves indicate four equal-thickness segments underneath the renal membrane (offset: 0.25~1.25 mm) to calculate the mean arterial or venous blood flow speeds of the four segments. The blood flow speed map of (b) arterial and (c) venous microvessels are displayed and quantified separately.

The renal membrane was manually depicted on the B-mode image and used in quantification of the microvasculature. A 1 mm range in the cortex was divided into four segments: 0.25 ~ 0.5 mm, 0.5 ~ 0.75 mm, 0.75 mm ~ 1 mm, and 1 ~ 1.25 mm underneath the membrane (0 mm) [Fig. 5(a)]. The speeds of MBs in each region were averaged to obtain the mean blood flow speeds of arteries and veins, and then the mean blood flow speeds of four segments were fitted to a straight line, whose slope was taken as the gradient of blood flow speed with depth (from proximal to distal microvessels). The superficial region (0 ~ 0.25 mm) was not included in the quantification because the number of MBs passing through this region was not enough to form a fully developed microvasculature for precise quantification. The upper bond (1.25 mm) of the range was determined to avoid the arcuate arteries in the quantification.

e) **Serum tests and pathological analysis**

After ultrasound imaging data acquisition, 3-4 ml blood was taken from the abdominal aorta or inferior cava. After centrifugation, the upper serum was taken to measure the concentration of serum creatinine, urea nitrogen, and β2 micro-globulin. Thereafter, the left kidney was taken out by laparotomy. The kidney specimens were dissected along the long axis, and fixed in 4% paraformaldehyde for pathological analysis. Renal tissues were paraffin-embedded and refrigerated. Then, the tissue specimens were consecutively sectioned with a thickness of about 2-3 μm in a microtome. Hematoxylin-eosin (H&E) staining of vascular endothelial cells were performed. The morphology and thickness of glomerular basement membrane and mesangial membrane were observed by a CX-21 biological microscopy (Olympus, Tokyo, Japan) with a magnification factor of 100.

3. Results

Statistical analysis was performed with Student's test for unpaired data. The 95% limit of probability was considered to be significant.

a) **Routine clinical indicators**

The measurements of blood pressure and heart rate in two groups of rats are shown in Fig. 6. It can be seen that both the systolic and diastolic pressures of the SHR group are significantly higher than those of the WKY group, and the heart rate does not show significant difference between the two groups.

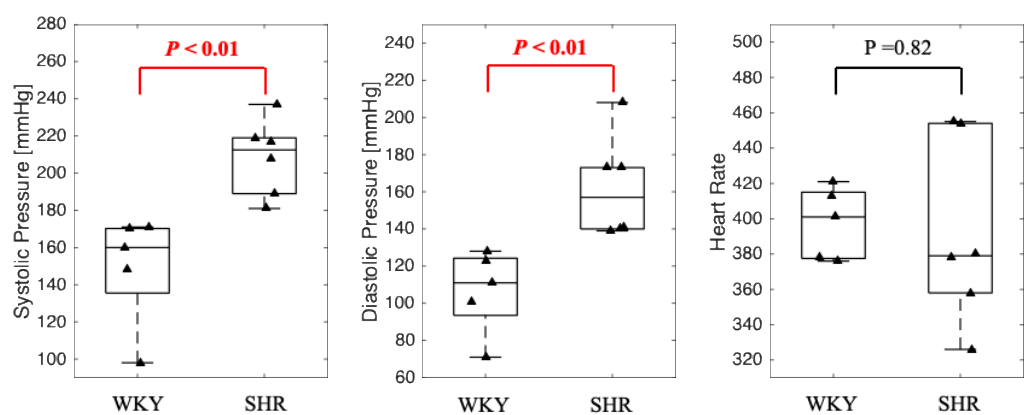

**Figure 6.** Measurements of (a) systolic pressure, (b) diastolic pressure and (c) heart rate, respectively.

Statistical results of serum detection indexes, including β2-microglobulin (p=0.74), serum urea nitrogen (p=0.71) and creatinine (p=0.91), do not show significant difference between the two groups. The results of 24-hour urine volume (p=0.95) and urine protein (p=0.82) in the two groups of rats do not show significant difference.

b) **Conventional ultrasound imaging**

Fig. 7 shows the Doppler ultrasound imaging parameters. The above three Doppler ultrasound quantitative parameters in the two groups do not show significant difference.

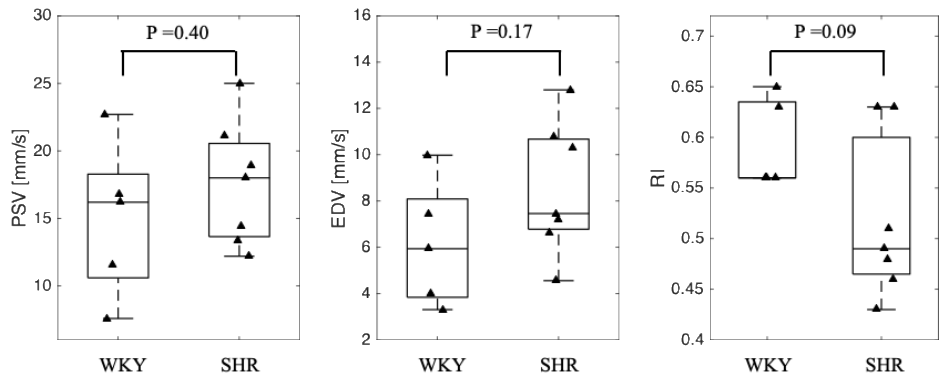

**Figure 7.** Doppler ultrasound imaging parameters: (a) PSV, (b) EDV, (c) RI.

The results of the five quantitative parameters from the CEUS perfusion curves in the two groups of rats are shown in Fig. 8. Four parameters (RT, IMAX, mTT and AUC) do not show significant difference between the two groups, while TTP shows a significant difference between the two groups.

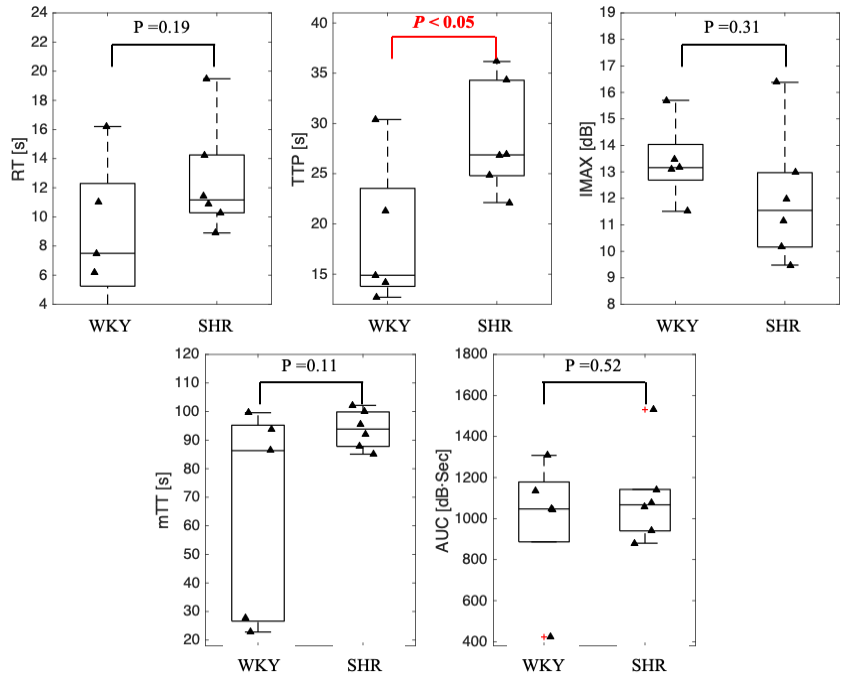

**Figure 8.** Quantitative parameters of CEUS perfusion curve. (a) RT, (b) TTP, (c) IMAX, (d) mTT, (e) AUC.

**c) Pathological sections**

The assessment of H&E stained renal pathological sections suggests a relatively mild renal damage, and some typical results are presented in Fig. 9. More specifically, no significant change is observed in most of the glomeruli, while only a few of histologic alterations including slight shrinkage of glomerular capillary loops [Fig. 9(b)] and hyaline casts [Fig. 9(c)] are observed. Slight intimal and media thickening, as well as the narrowing of corresponding arteriole lumina [Fig. 9((e) and (f))] can be observed.

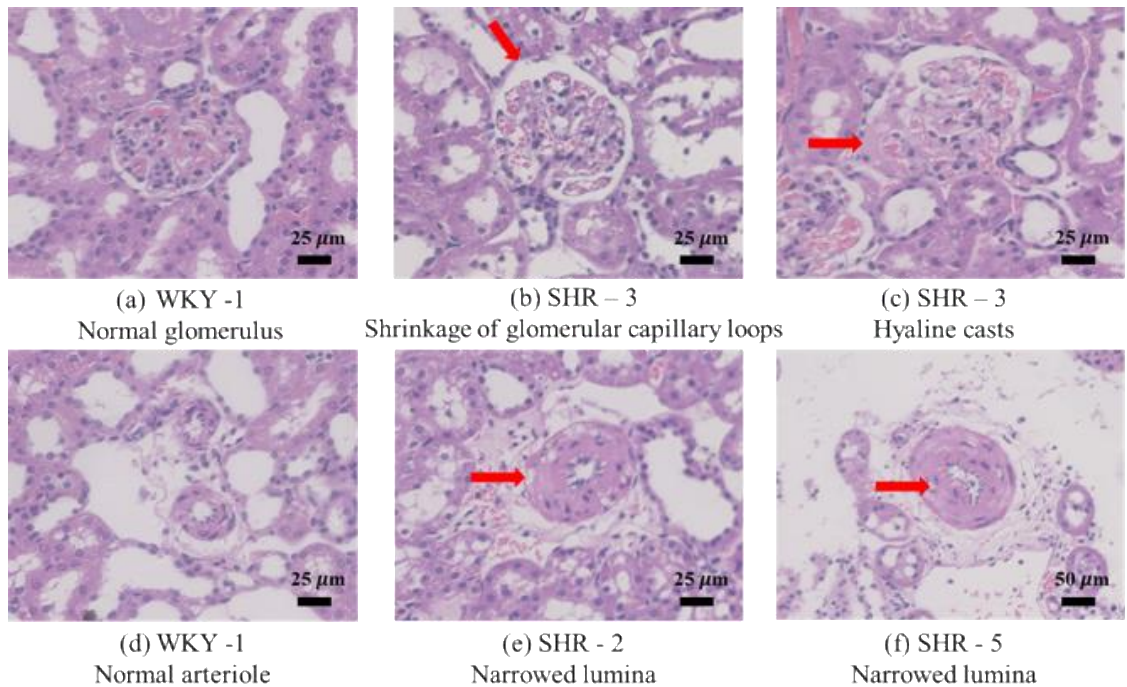

**Figure 9.** Typical results of H&E staining of pathological sections from WKY rats ((a) and (d)) and SHR rats ((b), (c), (e) and (f)), respectively. The red arrows indicate the shrinkage of glomerular capillary loops (b), hyaline casts (c) and narrowed lumina ((e) and (f)), respectively.

**d) ULM**

The results of ULM of the rat kidney are shown in Fig. 10. Fig. 10(a) is the blood flow direction map. Fig. 10(b) is the blood flow speed map, from which arterial blood flow speed [Fig. 10(c)] and venous blood flow speed [Fig. 10(d)] can be derived. It can be seen that the arterial blood flow speed [Fig. 10(c)] is higher than the adjacent venous blood flow speed [Fig. 10(d)].

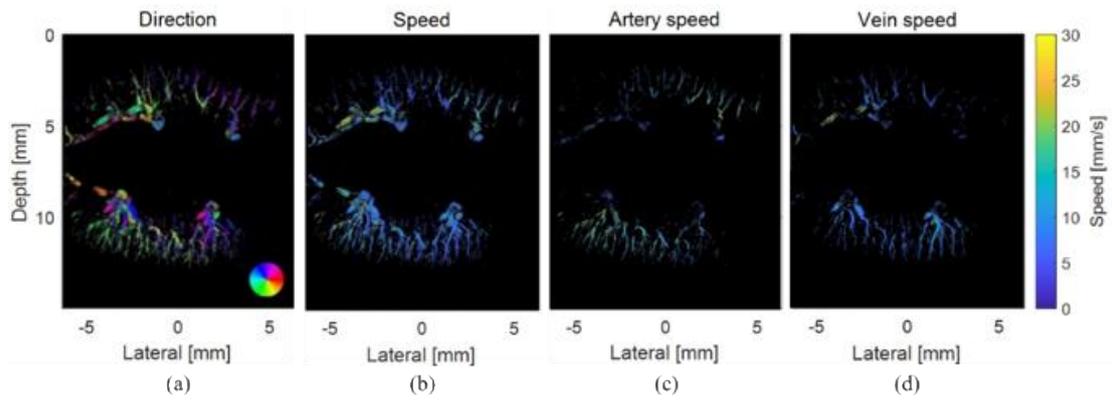

**Figure 10.** Maps of ULM from a rat kidney. (a) The blood flow direction map, with the color bar shown as the color disk on the lower right. (b-d) The blood flow speed maps of (b) artery and vein, (c) artery, and (d) vein, respectively, with the color bar shown on the right.

The mean arterial and venous blood flow speeds of different segments (as indicated in Fig. 5 (a)) and blood flow speed gradients are calculated, and statistical results are shown in Figs. 11, 12 and 13.

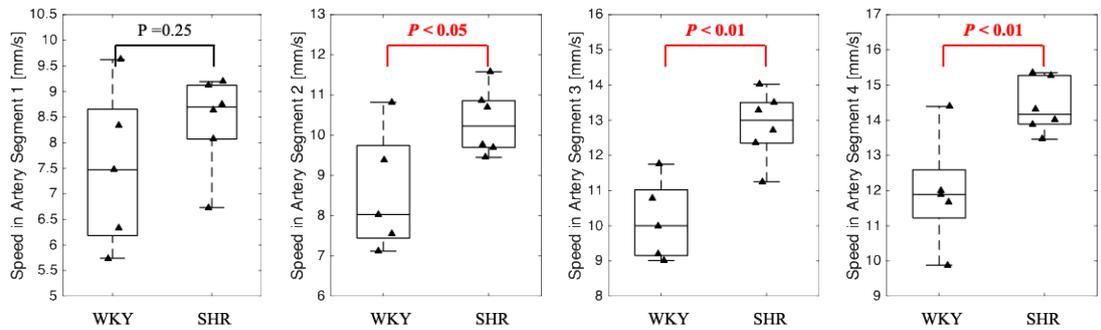

**Figure 11.** Mean arterial blood flow speeds of different segments for WKY and SHR groups. Segments 1 to 4 correspond to the four regions from the superficial region to juxtamedullary region.

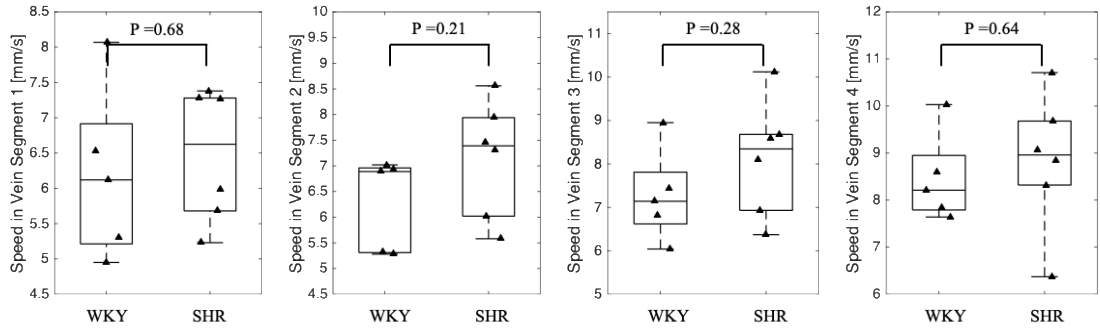

**Figure 12.** Mean venous blood flow speeds of different segments for WKY and SHR groups. Segments 1 to 4 correspond to the four regions from the superficial region to juxtamedullary region.

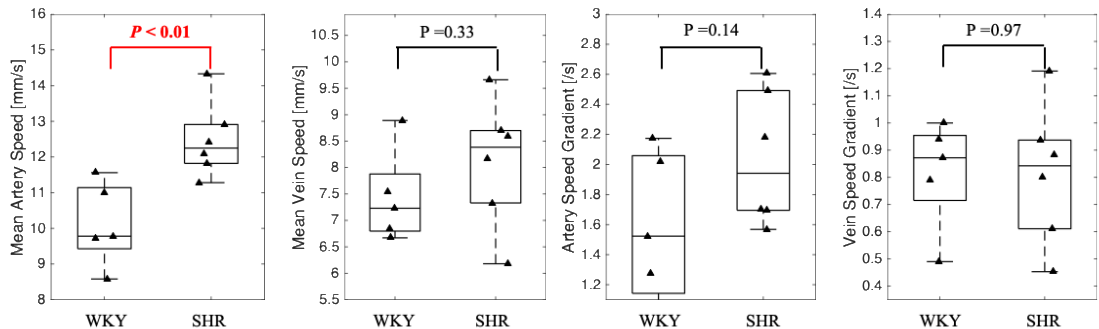

**Figure 13.** Mean arterial (a) and venous (c) blood flow speed of all segments for WKY and SHR groups. Gradient of arterial (b) and venous (d) blood flow speed from segments 1 to 4.

It can be seen that the mean arterial blood flow speed of segments 2, 3 and 4 and the mean arterial blood flow speed of all artery segments of SHR are significantly higher than those of WKY. A slight but non-significant increase of artery gradient is also observed. Parameters related to the venous blood flow speeds do not show significant difference between the two groups.

## 4. Discussion

In this study, assessment of hypertensive kidney disease using ULM was preliminarily demonstrated. Compared with conventional CEUS technique, ULM has rather higher spatial resolution to obtain more abundant hemodynamic information within the renal microvessels. Furthermore, the experimental results showed that the indicators obtained by conventional ultrasound imaging, urine and serum test have no statistical significance in the early stage of HN; on the contrary, several indexes related to the microvascular blood flow speed estimated using ULM generally has statistical significance.

Compared with that of the WKY rats, the blood pressure of the SHR usually rises during the first 10 weeks after birth and would be significantly higher at 4 weeks of age.[48] The measured significantly increased systolic pressure and diastolic pressure [Fig. 6] can verify the successful development of hypertension for the SHR group. It has been reported that high systolic blood pressure is the most related factor to hypertensive renal damage.[49,50] However, through the first 10 weeks, the autoregulatory efficiency would gradually increase to help to maintain normal intra-glomerular pressure and to protect glomeruli from damage caused by the increased blood pressure.[3] As a result, no significant change in renal histology could be observed in 2-month-old rats using optical microscopy.[51] Therefore, it may be difficult to detect HN at early stage using glomerular-damage-related or tubule-damage-related indicators. This assumption agrees with our experiments that urine volume, urine protein (late-stage indicator for significant urinary abnormality) [52] and other serum-detection-based indicators obtained from the SHR were not significantly different from those from the WKY group. Renal autoregulation would also help to maintain a stable RBF when the arterial pressure increases.[53] Therefore, for conventional ultrasound imaging, no statistically significant differences are observed between the Doppler-related parameters (PSV, EDV and RI) and CEUS-related parameters (RT, IMAX, mTT and AUC). A significant difference is observed between the CEUS-related parameter TTP. However, it should be noted that the analysis of CEUS TIC curve was performed using LDRW-WIWO model provided by QLAB, which is not completely in accordance with the rat kidney perfusion model. In summary, owing to renal autoregulation, the renal damage caused by hypertension at this stage is relatively mild and cannot be faithfully detected using the conventional clinical detection indexes, and the alteration of renal blood flow is relatively mild, especially in the large artery, to be detected by conventional ultrasound imaging methods.

Even with the protection of renal autoregulation, some renal damages (glomerulosclerosis, arteriolar hyalinosis, intimal and media thickening of arterioles) have already happened in the SHR, as shown in the H&E staining of pathological sections [Fig. 9]. Therefore, it is vital to develop an *in-vivo* high-resolution imaging tool to detect the microvascular changes in the early stage of HN. This study obtained the blood flow speed of microvessels by tracking the trajectory of MBs, and discovered that the mean blood flow speeds of the artery segments 2, 3 and 4 and the mean blood flow speed of all artery segments of the SHR are significantly higher than those of the WKY rats. A slight but non-significant increase of arterial blood flow speed gradient is also observed.

The findings agree with the results in the literature about the mechanisms of renal autoregulation. It should be noted that the quantitative results in this study are considered to reflect the blood flow speed changes of interlobular arteries of the two groups owing to the overlap between the defined quantification regions and the interlobular arteries. It has been reported that the interlobular artery participates in the adjustment of the renal resistance by myogenic mechanism to realize autoregulation.[54,55] A more specific explanation is that the dilatation and constriction of the interlobular arteries contribute to renal autoregulation.[56] Considering that the renal blood flow (RBF) is similar in the SHR and WKY groups at the ages from 2 to 10 weeks [48,57] and before the age of 15 weeks,[58] the constricted interlobular arteries may lead to acceleration of their blood flow, which agree with our findings of increased mean arterial blood flow speed. The stable RBF can be attributed to the increase of renal resistance when arterial pressure increases,[53] which would intuitively cause the increase of gradient of arterial blood flow speed as well. However, in this study, only a non-significant increase of the arterial blood flow speed gradient is observed. This is probably because the resistance is much greater in the inner part of the interlobular artery than in the outer part.[54] However, in this study, the interlobular arteries were divided into four segments and the gradient of their mean arterial blood flow speeds cannot completely reflect the greatly increased resistance in the inner part. Another interesting finding is that the mean arterial blood flow speeds of segments 2, 3 and 4 in the SHR are significantly higher than those in the WKY group, while no significant difference is observed in the mean arterial blood flow speed of segment 1 (superficial cortex) between the SHR and WKY groups. This is an indirect evidence of renal autoregulation of interlobular artery and is in accordance with the results in the literature that the glomerular pressures of the SHR and WKY groups are similar in the superficial glomeruli, but are significantly different in the juxtamedullary glomeruli.[59]

Parameters related to venous blood flow speed, including the mean venous blood flow speeds of each segment and mean venous blood flow speed of all segments, do not demonstrate statistically significant difference between the SHR and WKY groups. This is probably because that renal autoregulation is mainly achieved by the artery.

Despite the promising diagnosis capability of ULM for early HN indicated by the results, this study has several limitations. The microvasculature and the corresponding velocity maps are obtained using ULM by accumulating a large amount of MB events and tracking their trajectories. The passage rate of MBs through the vessels is highly related to the size of the vessels. Hingot et al., reported that large vessels of 100 micrometers could be fully reconstructed within 10 seconds, while the detection of capillaries would take up to tens of minute.[60] Because of the extremely large amount of data acquired using PW imaging, the data acquisition only lasted for 15 seconds in this study. Therefore, the vessel density and diameters of microvessels cannot be accurately quantified in this study and the quantification mainly relied on the blood flow speed measurements along the interlobular artery. The prolonged acquisition time would enable the fully reconstruction of smaller vasculature, including the afferent arterioles or capillaries of the glomerulus. It would enable a better study of *in-vivo* arteriole-based autoregulation [61] and possibly enable the finer diagnosis of damage level for different renal regions, such as juxtamedullary (the damage to this region is often considered to be more severe) or superficial regions.[62]

Conventional ultrasound imaging methods including Doppler and CEUS imaging were conducted using a clinical ultrasound scanner, which is not optimal for imaging the rats owing to its limited spatiotemporal resolution and other optimizations made for human subjects. An imaging platform with higher spatiotemporal resolution may improve the accuracy of Doppler and CEUS measurements, and the corresponding diagnosis.

## 5. Conclusion

In this study, experimental comparison was performed between 6 SHR and 5 normal WKY rats at age of 10 weeks to investigate the feasibility of ULM in diagnosing HN at the early stage. We performed pathology, urine, serum tests, CEUS and Doppler ultrasound imaging on the rats' kidneys. In addition, ULM was performed on the sagittal plane of the rats. The mean blood flow speeds and their gradients along depth of arteries and veins were statistically analyzed, respectively. The results demonstrate that the mean arterial blood flow speed has significant difference between the SHR and WKY groups, and can be used for early detection of HN. The advantages of ULM over conventional clinical inspection and ultrasound imaging methods for early diagnosis of hypertensive nephrosclerosis was validated.

**Disclosure statement**

The authors declare that they have no known competing financial interests or personal relationships that could have appeared to influence the work reported in this paper.

**Acknowledgements & Funding:**

This study was supported in part by the National Natural Science Foundation of China (61871251, 61801261 and 62027901), the China Postdoctoral Science Foundation (2017M620802), Sichuan Science and Technology Program (2019YFSY0048), Tsinghua-Peking Joint Center for Life Sciences, the Young Elite Scientists Sponsorship by China Association for Science and Technology and Beijing Municipal Administration of Hospitals' Ascent Plan (DFL20180102).

bibliography*Sci Rep*. 2017;7(1):1-12. doi:10.1038/s41598-017-13676-7

31. Zhu J, Rowland EM, Harput S, et al. 3D super-resolution US imaging of rabbit lymph node vasculature in vivo by using microbubbles. *Radiology*. 2019;291(3):642-650. doi:10.1148/radiol.2019182593

32. Yang Y, He Q, Zhang H, et al. Assessment of diabetic kidney disease using ultrasound localization microscopy: an in vivo feasibility study in rats. *IEEE Int Ultrason Symp IUS*. 2018:5-8.

33. Andersen SB, Hoyos CAV, Taghavi I, et al. Super-Resolution Ultrasound Imaging of Rat Kidneys before and after Ischemia-Reperfusion. *IEEE Int Ultrason Symp IUS*. 2019;2019-Octob:1169-1172. doi:10.1109/ULTSYM.2019.8926190

34. Chen Q, Yu J, Rush BM, Stocker SD, Tan RJ, Kim K. Ultrasound super-resolution imaging provides a noninvasive assessment of renal microvasculature changes during mouse acute kidney injury. *Kidney Int*. 2020;98(2):355-365. doi:10.1016/j.kint.2020.02.011

35. Dencks S, Piepenbrock M, Opacic T, et al. Clinical Pilot Application of Super-Resolution US Imaging in Breast Cancer. *IEEE Trans Ultrason Ferroelectr Freq Control*. 2019;66(3):517-526. doi:10.1109/TUFFC.2018.2872067

36. Okamoto K, Aoki K. Development of a strain of spontaneously hypertensive rats. *Jpn Circ J*. 1963;27:282-293. http://www.mendeley.com/research/geology-volcanic-history-eruptive-style-yakedake-volcano-group-central-japan/.

- 30 -